\documentclass[twocolumn,showpacs,preprintnumbers,amsmath,amssymb]{revtex4}

\usepackage{graphicx}
\usepackage{dcolumn}
\usepackage{bm}

\begin{document}

\preprint{APS/123-QED}

\title{Partial ferromagnetic ordering and indirect exchange interaction in spatially anisotropic kagome antiferromagnet Cs$_2$Cu$_3$CeF$_{12}$}

\author{T. Amemiya$^1$}
\author{M. Yano$^1$}
\author{K. Morita$^1$}
\author{I. Umegaki$^1$}
\author{T. Ono$^1$}
\author{H. Tanaka$^1$} 
\author{K. Fujii$^2$}
\author{H. Uekusa$^2$}
\affiliation{
$^1$Department of Physics, Tokyo Institute of Technology, Meguro-ku, Tokyo 152-8551, Japan\\
$^2$Department of Chemistry and Materials Science, Tokyo Institute of Technology, Meguro-ku, Tokyo 152-8551, Japan
}

\date{\today}

\begin{abstract}
We report the crystal structure and unconventional magnetic ordering of Cs$_2$Cu$_3$CeF$_{12}$, which is composed of buckled kagome lattice of Cu$^{2+}$ ions. The exchange network in the buckled kagome lattice is fairly anisotropic, so that the present spin system can be divided into two subsystems: alternating Heisenberg chains with strong antiferromagnetic exchange interactions and dangling spins. Although the direct exchange interactions between neighboring spins were found to be all antiferromagnetic, ferromagnetic magnetic ordering of the dangling spins was observed. Magnetization exhibits a plateau at one-third of the saturation magnetization. These observations can be understood in terms of the indirect interaction between dangling spins mediated by the chain spin.
\end{abstract}

\pacs{75.10.Jm; 75.40.Cx; 61.66.Fn}
\keywords{Cs$_2$Cu$_3$CeF$_{12}$, magnetic susceptibility, ferromagnetic ordering, magnetization plateau, specific heat, buckled kagome lattice, frustration}
\maketitle


Quantum magnetism of the Heisenberg kagome antiferromagnet (KAF) is of current interest from the viewpoint of the interplay between spin frustration and quantum many-body effects \cite{ML}. Although many theoretical studies on the Heisenberg KAF have concluded that the ground state for $S\,{=}\,1/2$ is disordered~\cite{Zeng1,Sachdev,Elstner,Nakamura}, its nature is still unresolved. Analytical and numerical studies for the $S\,{=}\,1/2$ case demonstrated that the ground state has a gap for triplet excitations but is gapless for singlet excitations \cite{Lecheminant,Waldtmann,Syromyatnikov,Jiang}. The magnitude of the triplet gap was estimated to be less than one-tenth of $J$ \cite{Waldtmann,Jiang,Singh2}. A valence-bond crystal represented by a periodic arrangement of singlet dimers~\cite{Singh2,Nikolic,Budnik,Yang} and a resonating valence bond state described by the linear combination of various configurations of singlet dimers~\cite{Mambrini,Hastings} have been proposed as the ground state of the $S\,{=}\,1/2$ Heisenberg KAF. Another theory based on a gapless critical spin liquid has also been proposed~\cite{Ryu,Hermele}. 

Experimental studies of the $S$=1/2 kagome antiferromagnet have been performed on Cu$_3$V$_2$O$_7$(OH)$_2$\,$\cdot$\,2H$_2$O \cite{Hiroi}, ZnCu$_3$(OH)$_6$Cl$_2$ \cite{Mendels,Helton,Imai,Olariu}, BaCu$_3$V$_2$O$_8$(OH)$_2$ \cite{Okamoto} and the A$_2$Cu$_3$MF$_{12}$ family with ${\rm A}={\rm Cs}$ and Rb, and ${\rm M}={\rm Zr}$, Hf and Sn \cite{Mueller,Morita,Ono}. A singlet ground state with a finite triplet gap was observed in Rb$_2$Cu$_3$SnF$_{12}$, which has a $2\,{\times}\,2$ enlarged chemical unit cell \cite{Morita,Ono}. 

Recently, the ground state for the spatially anisotropic KAF with $J\,{\neq}\,J'$ has been discussed theoretically \cite{Wang,Ya,Stoudenmire,Schnyder}, where $J$ and $J'$ are the exchange interactions along one direction and the other two directions on the kagome lattice, respectively. Some unusual ground states were predicted for the extremely anisotropic case of $J\,{\gg}\,J'$ \cite{Ya,Stoudenmire,Schnyder}. In general, it is considered that when the exchange interactions become spatially anisotropic, the spin frustration is reduced. However, such a spatially anisotropic model is useful for better understanding the isotropic case. It has been found in the $S$=1/2 triangular antiferromagnet Cs$_2$CuBr$_4$ \cite{Fortune} that spatial anisotropy in the exchange interactions enhances the frustration effect and leads to the cascade of quantum phase transitions in magnetic fields. 

In this paper, we report the crystal structure of the newly synthesized Cs$_2$Cu$_3$CeF$_{12}$ and its magnetic properties. In Cs$_2$Cu$_3$CeF$_{12}$, Cu$^{2+}$ ions form a kagome lattice, which is buckled like a staircase. From the anisotropic configuration of the hole orbitals of Cu$^{2+}$ ions, the exchange network on the buckled kagome lattice is largely anisotropic. Consequently, Cs$_2$Cu$_3$CeF$_{12}$ can be described as a spatially anisotropic $S$=1/2 Heisenberg KAF that is close to the case of $J\,{\gg}\,J'$.
To study the magnetic properties of Cs$_2$Cu$_3$CeF$_{12}$, we performed magnetization and specific heat measurements. As shown below, an unexpected ferromagnetic ordering of the dangling spins and a magnetization plateau at one-third of the saturation magnetization were observed, although the direct exchange interactions observed between neighboring spins were all antiferromagnetic. 


Cs$_2$Cu$_3$CeF$_{12}$ single crystals were grown from the melt of $\mathrm{CsF}$, $\mathrm{CuF_2}$ and $\mathrm{CeF_4}$. The materials were dehydrated by heating in vacuum at $\,{\sim}\,$100$^{\circ}$C and were packed into a Pt tube in the ratio of $3\,{:}\,3\,{:}\,2$. One end of the Pt tube was welded and the other end was tightly folded with pliers. Single crystals were grown from the melt. The temperature of the furnace was lowered from 750 to 500$^{\circ}$C over four days. Transparent light-blue crystals with a typical size of $2\,{\times}\,2\,{\times}\,0.5$ mm$^3$ and a platelet shape with a wide $(1,0,0)$ plane were produced.

We analyzed the structure of Cs$_2$Cu$_3$CeF$_{12}$ at room temperature using a Bruker SMART-1000 three-circle diffractometer equipped with a CCD area detector. Monochromatic Mo-K$\alpha$ radiation was used as an X-ray source. Data integration and global-cell refinements were performed using data in the range of $2.31^\circ\,{<}\,{\theta}\,{<}\,27.55^\circ$, and multiscan empirical absorption correction was also performed. The total number of reflections observed was 15191. 1464 reflections were found to be independent and 1374 reflections were determined to satisfy the criterion $I\,{>}\,2{\sigma}(I)$. Structural parameters were refined by the full-matrix least-squares method using SHELXL-97 software. The final $R$ indices obtained were $R\,{=}\,0.021$ and $wR\,{=}\,0.049$. 

Magnetization was measured in the temperature range of $1.8-400$ K using a SQUID magnetometer (Quantum Design MPMS XL). The specific heat was measured by the relaxation method in the temperature range of $1.8-300$ K using a Physical Property Measurement System (Quantum Design PPMS). 

\begin{figure}[t]
\includegraphics[width=7.3cm, clip]{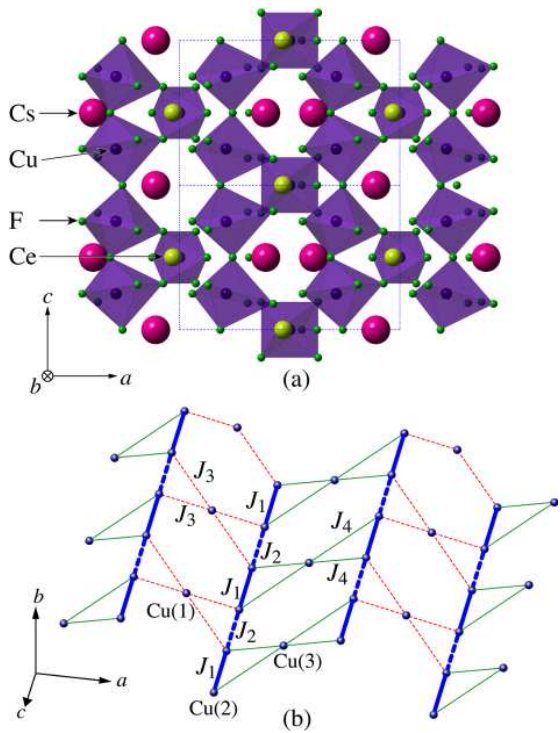}
\caption{(Color online) (a) Crystal structure of Cs$_2$Cu$_3$CeF$_{12}$ viewed along the $b$ axis. Shaded octahedra represent CuF$_6$ octahedra. Dotted lines denote the unit cell. (b) Exchange network in the buckled kagome lattice of Cu$^{2+}$ ions.}
 \label{fig:structure}
 \end{figure}


The crystal structure of Cs$_2$Cu$_3$CeF$_{12}$ is orthorhombic, $Pnnm$, with cell dimensions of $a\,{=}\,11.0970(16)$ \AA,  $b\,{=}\,14.441(2)$ \AA\, and $c\,{=}\,7.2970(11)$ \AA, and $Z\,{=}\,4$. Fractional atomic coordinates and equivalent isotropic displacement parameters are listed in Table~\ref{table:2}. Figure~\ref{fig:structure}(a) shows the crystal structure of Cs$_2$Cu$_3$CeF$_{12}$ viewed along the $b$ axis. All Cu$^{2+}$ ions are surrounded octahedrally by six F$^-$ ions. CuF$_6$ octahedra centered at Cu(2) and Cu(3) are elongated along one of the principal axes because of the Jahn-Teller effect, while those centered at Cu(1) are compressed. Consequently, the orbital ground state for Cu(1) is $d(3z^2\,{-}\,r^2)$. 
  
\begin{table}
\caption{Fractional atomic coordinates and equivalent isotropic displacement parameters ($\rm{\AA}^2$) for Cs$_2$Cu$_3$CeF$_{12}$.}
\label{table:2}
\begin{ruledtabular}
\begin{tabular}{lllll}
Atom  &  $x$  & $y$  & $z$ & $U_{\rm eq}$  \\ \hline
Cs(1) & 0.39152(3) & 0.37807(3) & 0.5 & 0.02229(10)   \\
Cs(2) & 0.65436(4) & 0.09909(3) & 0.5 & 0.02946(11)  \\
Cu(1) & 0.5 & 0.5 & 0 & 0.00989(17)  \\
Cu(2) & 0.71141(4) & 0.36371(3) & 0.24918(6) & 0.01419(11)  \\
Cu(3) & 0.5 & 0 & 0 & 0.0167(2)  \\
Ce & 0.47035(3) & 0.21758(2) & 0 & 0.01215(9)   \\
F(1) & 0.6246(3) & 0.1196(3) & 0  & 0.0343(10)  \\
F(2) & 0.4255(2) & 0.08202(16) & 0.1797(4) & 0.0246(5)  \\
F(3) & 0.5587(2) & 0.25899(18) & 0.2501(4) & 0.0239(5)   \\ 
F(4) & 0.4455(3) & 0.3702(2) & 0 & 0.0240(8)   \\ 
F(5) & 0.7412(6) & 0.3374(4) & 0 & 0.068(2)   \\ 
F(6) & 0.6291(3) & 0.47448(17) & 0.1922(3) & 0.0261(6)   \\ 
F(7) & 0.6847(3) & 0.3919(3) & 0.5 & 0.0231(8)   \\ 
F(8) & 0.8096(2) & 0.25820(15) & 0.3141(3) & 0.0167(5)   \\ 
\end{tabular}
\end{ruledtabular}
\end{table}

As shown in Fig.~\ref{fig:structure}(b), Cu$^{2+}$ ions with $S=1/2$ form a kagome lattice in the $ac$ plane, which is buckled with the appearance of a staircase. In Fig.~\ref{fig:structure}(b), solid lines denote exchange bonds. The hole orbitals $d(x^2\,{-}\,y^2)$ of Cu(2) ions are linked through $p$ orbitals of F(5) and F(7) ions along the $c$ axis with a Cu(2)${-}$F${-}$Cu(2) bond angle of approximately $149^{\circ}$. Thus, the exchange interaction between Cu(2) ions should be antiferromagnetic and strong, as observed in hexagonal A$_2$Cu$_3$MF$_{12}$ systems \cite{Morita,Ono}. Because the Cu(2)${-}$Cu(2) distance alternates slightly along the $c$ axis, the exchange interactions between Cu(2) ions should be alternating. We label the exchange interactions with Cu(2)${-}$F${-}$Cu(2) bond angles of 149.23$^{\circ}$ and 149.01$^{\circ}$ as $J_1$ and $J_2$, respectively. Since the difference between the bond angles is very small, the alternation parameter ${\alpha}\,{=}\,J_2/J_1$ should be close to unity. The exchange interactions $J_3$ between Cu(1) and Cu(2) and $J_4$ between Cu(2) and Cu(3) must be much weaker than $J_1$ and $J_2$, because the hole orbitals of these Cu$^{2+}$ ions are not directly linked through $p$ orbitals F(6) and F(2). Therefore, we can deduce that the exchange network of the kagome staircase in Cs$_2$Cu$_3$CeF$_{12}$ is decomposed into two subsystems: alternating-exchange chains running along the $c$ axis and dangling spins, which are weakly coupled to the chain spins through $J_3$ and $J_4$.

\begin{figure}[htbp]
\begin{center}
\hspace{-0.2cm}\includegraphics[width=7.8cm, clip]{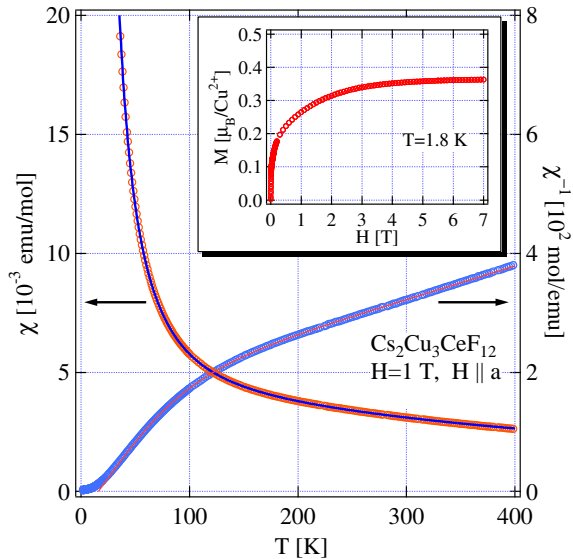}
\end{center}
\caption{(Color online) Temperature dependences of magnetic susceptibility and inverse magnetic susceptibility in Cs$_2$Cu$_3$CeF$_{12}$ measured for $H\,{\parallel}\,a$. Solid lines are the fits given by eqs.~(\ref{eq:sus1}) and (\ref{eq:sus2}) with the exchange parameters and $g$ factors shown in the text. The inset shows the magnetization curve measured at 1.8 K.}
 \label{fig:sus}
 \end{figure}

Figure \ref{fig:sus} shows the temperature dependences of magnetic susceptibility and inverse magnetic susceptibility in Cs$_2$Cu$_3$CeF$_{12}$ measured at 1.0 T for $H\,{\parallel}\,a$. 
The magnetic susceptibility increases monotonically with decreasing temperature, which is in contrast to the temperature dependence of other members of the A$_2$Cu$_3$MF$_{12}$ kagome family \cite{Morita,Ono}. The magnetic susceptibility above 200 K obeys the Curie-Weiss law with the Weiss constant ${\Theta}\,{\simeq}-250$ K. This large Weiss constant arises from the spin chains with strong antiferromagnetic exchange interactions $J_1$ and $J_2$. The rapid increase of the magnetic susceptibility below 100 K is attributed to the paramagnetic susceptibility of the dangling spins, which are weakly coupled to the chain spins through $J_3$ and $J_4$. 

\begin{figure}[htbp]
\begin{center}
\includegraphics[width=8.5cm, clip]{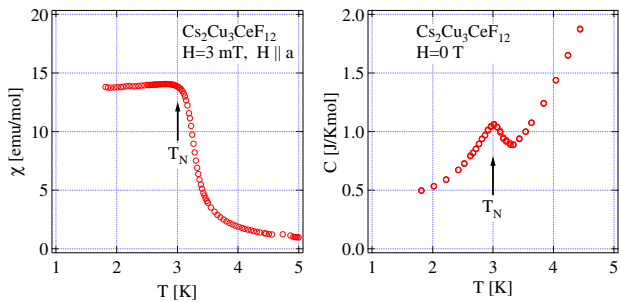}
\end{center}
\caption{(Color online) (a) Low-temperature magnetic susceptibility measured at 3.0 mT for $H\,{\parallel}\,a$. (b) Low-temperature specific heat measured at zero field. Arrows indicate the Curie temperature $T_{\rm C}$ determined from the specific heat measurement. }
\label{fig:lowsus}
\end{figure}

With further decreasing temperature, the magnetic susceptibility measured at a small magnetic field exhibits a sharp increase at $T\,{\sim}\,3$ K and becomes almost constant, as shown in Fig.~\ref{fig:lowsus}(a). This indicates that ferromagnetic ordering occurs at $T\,{\sim}\,3$ K. Figure~\ref{fig:lowsus}(b) shows the low-temperature specific heat measured at zero field. A cusplike anomaly, indicative of magnetic ordering, is observed at $T_{\rm C}\,{=}\,3.00(5)$ K. However, with increasing external field, the cusplike anomaly collapses rapidly. Such field dependence of the specific heat is typical of the ferromagnetic phase transition.
 
The inset of Fig.~\ref{fig:sus} shows magnetization curves measured at 1.8 K for $H\,{\parallel}\,a$. With increasing magnetic field, magnetization increases sharply and exhibits a plateau at one-third of the saturation magnetization. Because the numbers of chain spins and dangling spins are in the ratio of $2\,{:}\,1$, it is natural to consider that the 1/3 magnetization plateau arises from the full polarization of the weakly coupled dangling spins, and that the ordered moment of the chain spin is negligible. Note that this spin state is different from the {\it up-up-down} spin arrangement in the 1/3 magnetization plateau for the uniform kagome antiferromagnet \cite{Hida2}. From these observations, we can conclude that the magnetic phase transition at $T_{\rm C}\,{=}\,3.0$ K arises from the ferromagnetic long-range order of the dangling spins. The origin of the ferromagnetic exchange interaction between dangling spins is discussed later.
 
Next, to evaluate the individual exchange parameters, we analyze the magnetic susceptibility using the mean field approximation. Here, we define the exchange constant $J_{ij}$ as ${\cal H}_{\rm ex}\,{=}\,\sum_{\langle i,j\rangle} J_{ij}\,{\bm S}_i\,{\cdot}\,{\bm S}_j$. We write the magnetic 
susceptibilities of chain spins and dangling spins as ${\chi}_{\rm ch}$ and ${\chi}_{\rm d}$, respectively. When the $J_3$ and $J_4$ interactions vanish, ${\chi}_{\rm ch}$ is equivalent to the magnetic susceptibility of the $S=1/2$ alternating antiferromagnetic Heisenberg chain ${\chi}_{\rm ch}^0$, whose analytical form is given in the literature \cite{Johnston}. We assume that ${\chi}_{\rm d}$ obeys the Curie-Weiss law ${\chi}_{\rm d}^0\,{=}\,C/(T-{\Theta})$ with a small positive Weiss constant ${\Theta}$, because the dangling spins exhibit ferromagnetic ordering. When exchange interactions $J_3$ and $J_4$ are treated by the mean field approximation, the magnetic susceptibilities ${\chi}_{\rm ch}$ and ${\chi}_{\rm d}$ are expressed as
\begin{eqnarray}
\label{eq:sus1}
&&{\chi}_{\rm ch}={\chi}_{\rm ch}^0\frac{1-\displaystyle\frac{g_{\rm d}(J_3+J_4)}{4g_{\rm ch}k_{\rm B}(T-{\Theta})}}{1-\displaystyle\frac{3{\chi}_{\rm ch}^0}{2Ng_{\rm ch}^2{\mu}_{\rm B}^2}\frac{J_3^2+J_4^2}{k_{\rm B}(T-{\Theta})}}\,,\\
\, \nonumber \\
&&{\chi}_{\rm d}=\frac{Ng_{\rm d}^2{\mu}_{\rm B}^2}{12k_{\rm B}(T-{\Theta})}\left\{1-\frac{3(J_3+J_4)}{Ng_{\rm ch}g_{\rm d}{\mu}_{\rm B}^2}{\chi}_{\rm ch}\right\}\,,  
\label{eq:sus2}
\end{eqnarray}
where $N$ is the total number of spins. $g_{\rm ch}$ and $g_{\rm d}$ are the $g$ factors of the chain spins and dangling spins, respectively. Although the $g$ factors for Cu(1) and Cu(3) with dangling spins are actually different because of the different ligand fields, we assume for simplification that their $g$ factors are the same. From the 1/3 magnetization plateau shown in the inset of Fig.~\ref{fig:sus}, the value of $g_{\rm d}$ can be evaluated as $g_{\rm d}\,{=}\,2.18(2)$. The total magnetic susceptibility ${\chi}$ is given by ${\chi}={\chi}_{\rm ch}+{\chi}_{\rm d}$. Equations~(\ref{eq:sus1}) and (\ref{eq:sus2}) are incorrect near $T_{\rm C}$, because in general, ${\Theta}$ is larger than ${T_{\rm C}}$ owing to the development of the spin correlation.
Fitting this theoretical ${\chi}$ to the experimental susceptibilities for $50\,{\leq}\,T\,{\leq}\,350$ K, we obtain $J_1/k_{\rm B}\,{=}\,316(9)$ K, $J_2/k_{\rm B}\,{=}\,297(9)$ K, $J_{3,4}/k_{\rm B}\,{=}\,88(4)$ K, $J_{4,3}/k_{\rm B}\,{=}\,85(4)$ K, ${\Theta}\,{=}\,13(1)$ K and $g_{\rm ch}\,{=}\,2.47(2)$. 
Since $J_3$ and $J_4$ interactions are symmetric in eqs.~(\ref{eq:sus1}) and (\ref{eq:sus2}), we cannot determine which is larger in the present analysis. The solid lines in Fig.~\ref{fig:sus} show the susceptibility and inverse susceptibility calculated with these parameters. All the exchange interactions are antiferromagnetic. As expected from the crystal structure, the obtained exchange constants $J_1$ and $J_2$ are much larger than $J_3$ and $J_4$ and are close to each other, i.e., ${\alpha}\,{=}\,0.940$. From the Weiss constant ${\Theta}$, we see that the magnitude of the exchange interaction between dangling spins is of the order of $(J')^2/(k_{\rm B}J)\,{\simeq}\,24$ K, where $J\,{=}\,(J_1+J_2)/2$ and $J'\,{=}\,(J_3+J_4)/2$. 

The ground state for the case $J_1\,{=}\,J_2\,{=}\,J$, $J_3\,{=}\,J_4\,{=}\,J'$ and $J\,{\gg}\,J'$ was discussed theoretically by Schnyder {\it et al.} \cite{Schnyder}. They argued that the ferromagnetic and antiferromagnetic indirect exchange interactions ${\cal J}_1$ and ${\cal J}_2$ mediated by the chain spins act between neighboring dangling spins located on both sides of the spin chain and between those located along the spin chain, respectively. They showed that the magnitudes of ${\cal J}_1$ and ${\cal J}_2$ are of the order of $(J')^2/J$ and that ${\cal J}_2/|{\cal J}_1|\,{\simeq}\,0.70$. They demonstrated that owing to these indirect interactions, the dangling spins and chain spins form a spiral structure with static moments of the order of ${\mu}_{\rm B}$ and $(J'/J){\mu}_{\rm B}$, respectively. 

This ordering scenario should be applicable to the present system. In the exchange network shown in Fig.~\ref{fig:structure}(b), ${\cal J}_1$ and ${\cal J}_2$ correspond to the interactions between Cu(1) and Cu(3) and between neighboring Cu(1) or neighboring Cu(3) along the $c$ axis, respectively. If the condition ${\cal J}_2/|{\cal J}_1|\,{<}\,1/2$ is satisfied, the ferromagnetic ordering of the dangling spins can occur. Because of the alternating spin chain with a finite triplet gap in Cs$_2$Cu$_3$CeF$_{12}$, the spin correlation in the spin chain decreases more rapidly than in a uniform chain with increasing distance between spins. Thus, the above condition may be realized in Cs$_2$Cu$_3$CeF$_{12}$. We infer that the ferromagnetic ordering of dangling spins and the positive Weiss constant ${\Theta}$ in eqs.~(\ref{eq:sus1}) and (\ref{eq:sus2}) are caused by the indirect ferromagnetic exchange interaction ${\cal J}_1$ mediated by the chain spins, because ${\Theta}$ is of the order of $(J')^2/(k_{\rm B}J)$ and the direct exchange interaction between dangling spins should be antiferromagnetic, if any. $T_{\rm C}$ is one-eighth of $(J')^2/(k_{\rm B}J)$. This can be ascribed to the fact that the interlayer interaction is much smaller than ${\cal J}_1$.

In conclusion, we have reported the crystal structure and magnetic properties of Cs$_2$Cu$_3$CeF$_{12}$. This compound is composed of anisotropic kagome layer of Cu$^{2+}$ ions, which can be subdivided into alternating-exchange chains and dangling spins. The observed magnetic phase transition accompanied with a net moment and the 1/3 magnetization plateau can be attributed to the ferromagnetic ordering and the full polarization of the dangling spins, respectively. From the analysis of the magnetic susceptibility, we have evaluated individual direct exchange interactions, which were found to be all antiferromagnetic. From these results, we can deduce that the magnetic ordering of the dangling spins is stabilized by the indirect ferromagnetic exchange interaction mediated by the chain spin. Because the triplet gap for the alternating-exchange chain closes in a high magnetic field, the 1/3 plateau state becomes unstable. Thus, we expect that a quantum phase transition occurs with increasing magnetic field. This transition should be observed with the high-field magnetization measurement. A ferrimagnetic structure stabilized by quantum fluctuations is also expected above the critical field \cite{Wang,Stoudenmire}.
   
This work was supported by a Grant-in-Aid for Scientific Research (A) from the Japan Society for the Promotion of Science (JSPS) and a Global COE Program ``Nanoscience and Quantum Physics'' at Tokyo Institute of Technology funded by the Japanese Ministry of Education, Culture, Sports, Science and Technology. T.O. was supported by a Grant-in-Aid for Young Scientists (B) from JSPS.


\end{document}